# Long-lived quantum correlation by cavity-mediated subradiance

Kyu-Young Kim[1], Jin Hee Lee[1], Woong Bae Jeon[1], Dong Hyun Park[1], Suk In Park[2], Jin Dong Song[2], Changhyoup Lee[3]*, and Je-Hyung Kim[1]*

[1]Department of Physics, Ulsan National Institute of Science and Technology; Ulsan, 44919, Republic of Korea

[2]Center for Opto-Electronic Materials and Devices Research, Korea Institute of Science and Technology; Seoul, 02792, Republic of Korea

[3]Korea Research Institute of Standards and Science; Dajeon, 34113, Republic of Korea

*Corresponding author. Email: changhyoup.lee@gmail.com and jehyungkim@unist.ac.kr

**Abstract:**

Cooperative effects such as super(sub)radiance in quantum systems arise from the interplay among quantum emitters. While bright superradiant states have been extensively studied and yielded significant insights into cooperative phenomena, subradiant states have remained less explored due to their inherently dark state nature. However, subradiance holds significant potential as valuable quantum resources that exploit long-lived and large-scale entanglement, which is a key for advancing quantum information technologies. Here, we demonstrate a long-lived subradiant state among multiple quantum emitters coupled to a directional low $Q$ cavity. In a tailored photonic environment with balanced cavity dissipation, emitter-field coupling strength, and incoherent pumping, two coupled quantum dots exhibit a steady-state population in a subradiant state with highly negative cooperativity. As an important hallmark of a subradiant state, the system shows large photon bunching ($g^{(2)}(0) \gg 2$) and suppressed single-photon decay. In addition, controlling the excitation wavelength provides a useful tool for manipulating dephasing and the number of coupled emitters, which leads to significant changes in photon statistics. Our approach to inducing cavity-mediated subradiance paves the way for creating and harnessing quantum correlations in quantum emitters via a long-lived entangled quantum state, essential for quantum storage and metrology.

## Introduction

Non-interacting quantum emitters can be radiatively coupled to each other, resulting in strongly enhanced collective decay rates, so-called Dicke superradiance [1]. This phenomenon arises across very different scales from a few quantum emitters [2-4] to high-density ensembles [5-7] and has been demonstrated in a variety of quantum platforms such as atoms [8,9], superconductors [10], and semiconductors [2,11]. However, achieving photon-mediated collective interaction requires strict conditions such as subwavelength distance and spectral identity among the emitters. The limitation of short-range coupling can be overcome by introducing one-dimensional photonic structures such as optical fibers [12] or photonic waveguides [3,8,11], which extend the spatial range of radiative coupling. Nonetheless, the superradiant state, by its enhanced radiative nature, only lives short and disappears very quickly with a radiative decay process, which limits its usefulness for quantum applications, although exhibiting rich physics and interesting features.

Besides the Dicke superradiance, a collective interaction of quantum emitters may also lead to another phenomenon called subradiance, showing contrary behaviours to superradiance. Exploiting the subradiance has an important advantage of long-lived quantum entanglement [13,14], providing valuable resources for quantum information storage [15,16] and quantum metrologies [17]. Furthermore, the subradiant state is more robust to dephasing [17-20] and accesses a decoherence-free subspace that is useful for quantum computation [21,22]. Thanks to these features, it has been attracting intensive interest from the community [23,24] and demonstrated with atomic ensembles [16,25,26], molecules [4,27], and quantum dots [28]. However, the cooperativity among the emitters remains small due to large photonic dissipation in free space or waveguides. Therefore, the previous studies managed the superradiance and subradiance transition by precisely controlling the phase of a driving field, adjusting frequency detunings among emitters, or utilizing different decay rates between subradiant and superradiant states in a pulsed regime, rather than achieving steady-state subradiance, which is crucial for its continuous population in a strong collective state. Moreover, turning off the repumping laser leaves the system in a long-lived entangled state. Despite its importance and potential, achieving the steady-state subradiance remains challenging, and the unique feature of consequent giant photon bunching has only been theoretically predicted [14,29-31].

Here, we experimentally demonstrate a long-lived entangled state of two InAs quantum dots (QDs) in a steady-state regime by addressing cavity-mediated subradiance. A nanophotonic cavity with a low *Q* and highly directional emission plays a crucial role in opening a viable path to tailor the cooperativity of coupled multiple QDs. Under the condition of the balanced emitter-cavity coupling strength and cavity dissipative loss in a weak incoherent pumping regime, the QDs exhibit a dominant steady-state population in the subradiant state. The emergence of a long-lived entangled state could be probed in a non-destructive way by photon statistics measurement, exhibiting giant photon bunching ($g^{(2)}(0) > 8$), much exceeding the classical limit. A highly suppressed single-photon decay is also observed from slow antibunching decay curves. Furthermore, we demonstrate that the excitation wavelength can manipulate the cooperativity strength and the number of coupled QDs in a cavity, leading to a stark difference in their photon statistics. Our work, therefore, explores the potential for the generation of long-lived quantum correlations among multiple quantum emitters and the production of highly cooperative emissions.

**Results**

Figures 1a-c compare two two-level emitters with different photonic environments under continuous incoherent excitation. For uncoupled emitters whose distance is much larger than their wavelength ($d \gg \lambda$) in free space, the emitters do not couple to each other and individually emit photons with a radiative lifetime ($\Gamma_r$) into free space without any correlation (Fig. 1a). The photon statistic of independently emitted photons follows $g_{uc}^{(2)}(\tau = 0) = 1 - 1/N$ at the zero-time delay, showing that an antibunching effect decreases with the number of emitters ($N$). When two emitters are within their wavelength ($d < \lambda$) in free space, they are radiatively coupled, so no longer separable but able to form asymmetric (subradiant) and symmetric (superradiant) entangled states, so-called Dicke states, as shown in Fig. 1b. Incoherent pumping of those emitters predominantly leads to an occupation of a fully excited state ($|ee\rangle$), followed by transitions to the ground state ($|gg\rangle$) via a superradiant state ($|+\rangle = (|eg\rangle + |ge\rangle)/\sqrt{2}$). Such a cascade two-photon process with an enhanced decay rate ($2\Gamma_r$) increases two-photon coincident events and gives rise to a small bunching at zero-time delay ($g_{sup}^{(2)}(0) > g_{uc}^{(2)}(0)$). As more emitters are involved, a faster collective decay and a higher bunching signal of $g^{(2)}(0) > 1$ are observed [32], but the bunching peak is still limited by the classical limit ($g^{(2)}(0) = 2$). Interestingly, such a bunching behavior can be largely enhanced if the emitters are in a subradiant state ($|-\rangle = (|eg\rangle - |ge\rangle)/\sqrt{2}$). The subradiant state is inherently dark with a small decay rate, so a single-photon transition from $|-\rangle$ to $|gg\rangle$ is significantly suppressed, while the incoherent continuous excitation to the emitters in a non-decaying state opens an excitation path from $|-\rangle$ to $|ee\rangle$ and allows them to participate in the cascade two-photon decay process in a more deterministic way. Therefore, together with a suppressed single-photon decay channel in the subradiant state, the enhanced two-photon decay process by incoherent pumping leads to strongly correlated photon-pair generation and giant photon bunching much exceeding $g_{sub}^{(2)}(0) \gg 2$, as described in Fig. 1c. Therefore, such a large photon bunching provides an important signature of the emergence of a steady-state subradiance [14,29-31].

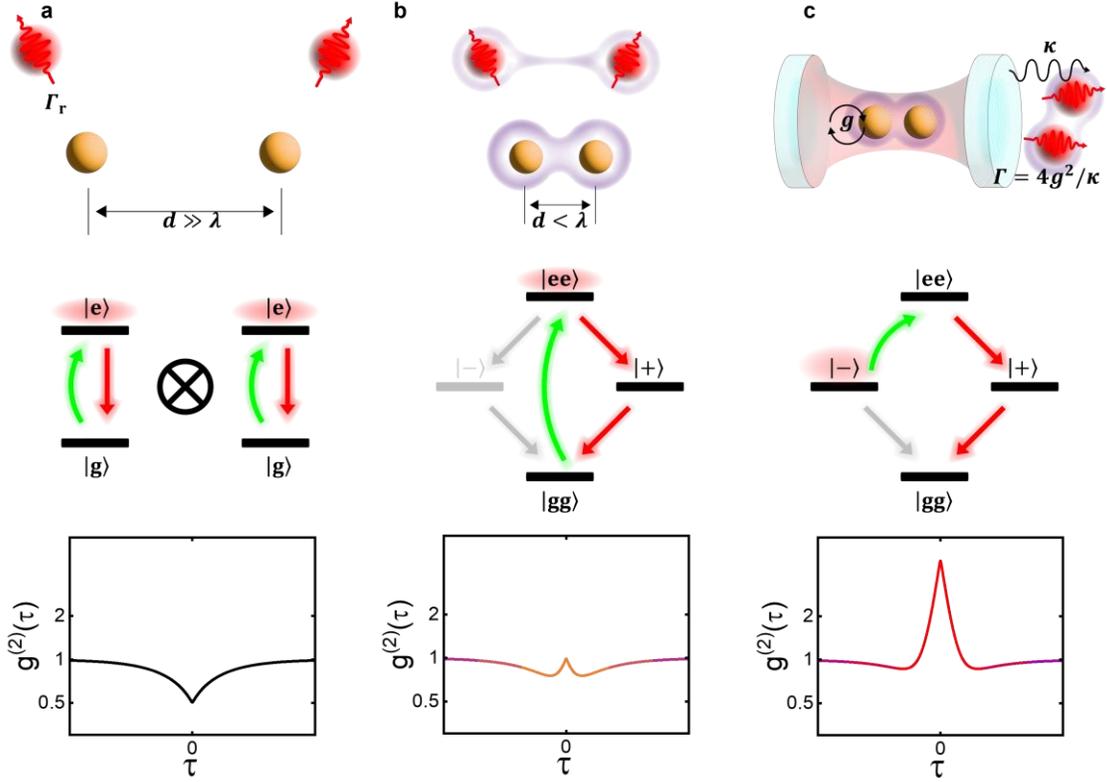

**Fig. 1. Comparison of collective effects in two quantum emitters.** We compare the schematics with different photonic environments (top), transitions via incoherent pumping and decay processes (middle), and photon statistics of two quantum emitters (bottom). **a**, Two uncoupled two-level emitters positioned more than their wavelength apart in free space ($d \gg \lambda$). Each emitter behaves independently and emits single photons with a spontaneous emission rate of $\Gamma_r$. The photon statistic yields a second-order coherence function, $g^{(2)}(0) = 0.5$, indicating independent photon emission. **b**, Two radiatively coupled quantum emitters separated by less than the wavelength at free space ($d < \lambda$). They form superradiant ($|+\rangle$) and subradiant ($|-\rangle$) states, and the decay from the fully excited state $|ee\rangle$ to the ground state $|gg\rangle$ occurs through the superradiant state $|+\rangle$ with an enhanced decay rate of $2\Gamma_r$. This configuration results in a mild bunching peak ($g^{(2)}_{\text{sup}}(0) > g^{(2)}_{\text{uc}}(0)$), reflecting the collective nature of the decay. **c**, Two coupled quantum emitters in a cavity. A balanced cavity dissipation and coupling strength enable a steady-state population in the subradiant state $|-\rangle$. This arrangement leads to superbunching ($g^{(2)}(0) \gg 2$) due to suppressed single-photon decay but enhanced two-photon decay with a collective decay rate $\Gamma = 4g^2/\kappa$.

**Steady-state subradiance by engineering cavity coupling**

Steady-state subradiance requires emitters to be coupled to a low $Q$ cavity, where the emitter-cavity coupling strength ($g$) and cavity's dissipative loss ($\kappa$) are balanced [14,23,29]. Therefore, the pumping and relaxation dynamics should be properly engineered through a tailored photonic environment to increase the steady-state population in a subradiant state. In this work, we demonstrate this from two-coupled QDs embedded in a highly directional and low $Q$ cavity, and the emergence of subradiance is monitored by their photon statistics, as described in Fig. 2a. Among several micro/nanocavities, we adopt a recently developed hole-based circular Bragg grating (hole-CBG) structure [33]. This cavity is optimized to produce a Gaussian far-field profile with very small divergence, ensuring that most of the emission is confined within a 5-degree angle, as shown in the top inset of Fig. 2a (Methods). Directional emission not only enhances a collection efficiency but also leads to high selectivity in momentum space. This plays an important role in enhancing cooperativity by spatially coupling the emitters and effectively removing which-path information when the emitted photons are detected [2,34]. Also, importantly, we designed the cavity to exhibit a simulated $Q$ value of 1,900, ensuring that its dissipation rate closely matches the coupling strength with emitters. The cavity was fabricated on InAs self-assembled quantum dots (QDs) embedded in a 160 nm GaAs membrane. The bottom inset of Fig. 2a shows a scanning electron microscope image of the fabricated hole-CBG overlapped with a calculated near-field image of the cavity mode.

First of all, we optically characterized cavity-coupled QDs using a 790 nm continuous wave laser at 7 K. The fabricated cavity shows a measured $Q$ value of 592, corresponding to $\kappa = 553 \pm 29.7$ GHz. For a single QD case, we observed a clear antibunching signal with $g^{(2)}(0) = 0.089 \pm 0.006$ (Supplementary Section 1). To investigate the cooperative effect further, we identified two resonant QDs at $\lambda_{\text{QDs}} = 916.18$ nm with a linewidth of $14.9 \pm 0.36$ GHz, as shown in Fig. 2b. These two QDs are spectrally overlapped within a spectral resolution of 7 GHz. Since cavity coupling is a crucial factor for enhancing collective cooperativity, we adjusted the spectral detuning ($\delta$) between the QDs and the cavity mode using a gas tuning technique [35]. Figures 2b-d compare the photoluminescence (PL) spectrum and photon correlation measurement data of cavity-coupled QDs with decreasing $\delta$ of 320, 194, and $-13.4$ GHz, respectively. At $\delta = 320$ GHz, the photon correlation curve exhibits an antibunching curve with $g^{(2)}(0) \sim 0.5$, indicative of two independent QDs in a spectrally detuned cavity. Noticeably, a small bunching peak appears at zero-time delay. As $\delta$ decreases to 194 GHz, a bunching peak increases, and when two QDs are spectrally well-coupled to the cavity ($\delta = -13.4$ GHz), a bunching peak exceeds the classical limit of $g^{(2)}(0) = 2$. This is an important indicator of the emergence of steady-state subradiance. The presence of both antibunching and bunching characteristics signifies single-photon and two-photon decay channels, respectively, discussed in Fig. 1c. To quantify the correlation data showing both bunching and antibunching features, we fit the experimental data using the following equation:

$$g^{(2)}(\tau) = \left[1 - (1-A)\exp\left(-\frac{|\tau|}{T_a}\right)\right]\left[1 + B\exp\left(-\frac{|\tau|}{T_b}\right)\right] \quad (1)$$

, where A (B) is a parameter related to an antibunching (bunching) amplitude with a time scale $T_a$ ($T_b$). In Fig. 2d, a nonclassical bunching signal reaches $g^{(2)}(0) = 3.27 \pm 0.07$ and shows $T_b = 61.6 \pm 2.66$ ps, corresponding to a collective decay rate $\Gamma = 4g^2/\kappa$. From this relation, we calculate the coupling strength $g = 47.4 \pm 1.6$ GHz. Thus, our cavity-coupled quantum emitters have a ratio of $\kappa/g \approx 11.7$. We note that a single-photon antibunching decay (yellow region) is also significantly influenced by the cavity detuning. At large cavity detuning ($\delta = 320$ GHz), $T_a$ is $17.1 \pm 0.22$ ns, but it extends to $29.6 \pm 0.23$ ns for the QDs with small $\delta = -13.4$ GHz. This extended single-photon decay time with small $\delta$ is unusual given that an improved spectral coupling typically shortens the recombination time through the Purcell effect, as we observed a very rapid recombination time of $121 \pm 0.92$ ps for the cavity-coupled QDs (Supplementary Section 1). These large photon bunching signal and elongated antibunching decay time clearly present the generation of a steady-state subradiant state that suppresses single-photon decay from $|-\rangle$ to $|gg\rangle$, and therefore, incoherent excitation to $|ee\rangle$ followed by two-photon pair emission becomes the only possible decay channel, as observed in Fig. 2d.

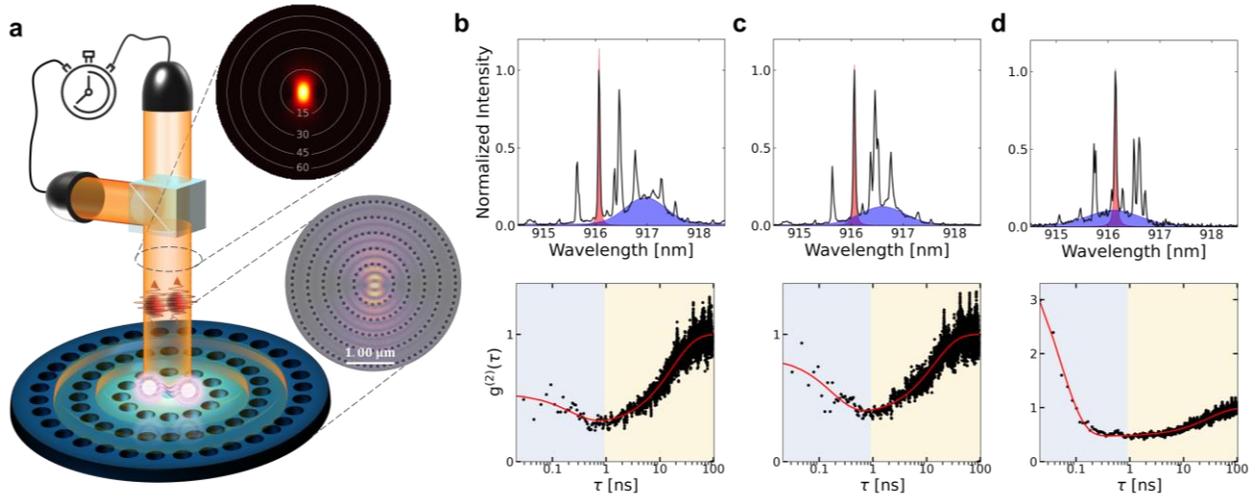

**Fig. 2. Cavity coupled QDs and detuning. a,** A schematic of an air-suspended hole-CBG cavity in InAs/GaAs membrane and photon correlation measurement. The upper inset is a calculated far-field image at the center frequency of the cavity mode, demonstrating a high directionality. The lower inset features a scanning electron microscope image of the hole-CBG sample, overlaid with a near-field profile of the cavity mode. **b-d,** Normalized spectra (top) and photon correlation curves (down) for two cavity-coupled QDs using a 790 nm continuous wave laser at various cavity detunings ($\delta$) of 320 (**b**), 194 (**c**), and $-13.4$ GHz (**d**), respectively. In the spectra, the cavity mode and the spectrally matched two QDs are fitted with Gaussian (blue) and Lorentzian (red) curves, respectively. Blue and yellow shaded regions in $g^{(2)}(\tau)$ denote bunching and anti-bunching profiles, respectively.

**Theoretical model of steady-state subradiance**

To quantitatively describe a steady-state collective system and its behaviors, we begin with the Tavis-Cummings Hamiltonian, including multiple cavity-coupled emitters. In a rotating frame, the Hamiltonian can be written as

$$H = \sum_{i=1}^{N} [\delta_i \sigma_i^+ \sigma_i^- + g(\sigma_i^+ a + \sigma^- a^\dagger)] \quad (2)$$

, where an identical coupling strength $g$ between individual emitters and a cavity mode is assumed, $a$ ($a^\dagger$) is an annihilation (creation) operator of the cavity mode, $\sigma_i^- = |g\rangle_i \langle e|_i$ defines a lowering operator of the $i$'th emitter, and $\delta_i$ is the frequency detuning between the cavity mode $\omega_c$ and the $i$'th emitter $\omega_i$. Along with the coherent interaction between the cavity mode and the emitters, we also take into account incoherent excitation and cavity dissipation loss. The time evolution of this coupled system, including such incoherent processes, can be expressed by the following Lindblad master equation:

$$\dot{\rho} = \mathcal{L}\rho = -i[H, \rho] + \kappa L[a] + \sum_{i=1}^{N} (PL[\sigma_i^+] + \gamma L[z_i] + \Gamma_r L[\sigma_i^-]) \quad (3)$$

and

$$L[A] = A\rho A^\dagger - \frac{1}{2}(A^\dagger A \rho + \rho A^\dagger A) \quad (4)$$

, where $\kappa$ is a dissipative cavity loss rate, $P$ is an incoherent pumping rate, $\gamma$ is a pure dephasing rate of the emitters, $\Gamma_r$ is a spontaneous emission rate of the emitters into other loss channels, not into the cavity mode, and $z_i = \sigma_i^\dagger \sigma_i$. We assume all the emitters are ideal and have a resonant frequency. In our model, we ignore the spontaneous decay as its rate is negligibly small compared to the other rates. Also, we disregard pure dephasing in quantum emitters to consider ideal quantum emitters, but we later introduce a dephasing term to account for experimental results.

From the simulated time-dependent solution $\rho(\tau)$ of the cavity-coupled two-level emitters when $N = 2$, we first obtain the steady-state density matrix $\rho_{ss}$ and investigate the steady-state population in each state: $\rho_{gg}$, $\rho_{ee}$, $\rho_+$, and $\rho_-$ (Supplementary Section 2). We analyze how these populations vary with the independent dissipative and pumping terms of $\kappa$ and $P$ in units of $g$ as shown in Figs. 3a-d. Noticeably, in a weak excitation regime ($P \ll g, \kappa$), $\rho_{gg}$ and $\rho_-$ are almost equally dominant. In this regime, the interplay between the fast collective decay of the radiatively coupled states and the slow, incoherent pumping accumulates the population dominantly in the subradiant state, striking a balance between $\rho_{gg}$ and $\rho_-$. As the pumping rate increases, $\rho_+$ starts to surpass $\rho_{gg}$ and $\rho_-$, entering into a superradiant regime. If the pumping rate further increases to a strong excitation regime ($P \gg g, \kappa$), the excitation rate to the fully excited state exceeds $\Gamma$, so $\rho_{ee}$ quickly becomes dominant. Therefore, it is clear from the numerical simulation that the steady-state subradiance is usually achievable in a very weak pumping regime. However, Fig. 3d shows that this subradiance dominant regime can be effectively expanded by adjusting $\kappa$. In particular, the subradiant state region enlarges when balanced with $g$ (maximum at $\kappa/g \approx 3$) and shrinks in the limit of low or high $\kappa$ (Supplementary Section 3). At high $\kappa$ ($\kappa/g \gg 1$), a significant cavity dissipation becomes dominant over the cavity-mediated coupling, leading to independent behaviors of the individual emitters, and consequently, the steady-state population quickly converges to the fully excited state. On the other hand, at low $\kappa$ ($\kappa/g \ll 1$), individual emitter-cavity coupling becomes stronger than collective coupling, and therefore, the emitters again tend to relax via independent channels rather than collective decay channels [29]. This clearly reveals that the interplay among the balanced dissipation, coupling strength, and incoherent pumping is crucial for fostering collective coherence and achieving steady-state subradiance.

The population in collective states yields cooperativity among the emitters, consequently affecting the photon statistics of the emitted photons, as explained below. Figure 3e illustrates the population difference in these collective states. When the steady-state population is dominated by super(sub)radiant states, the system exhibits positive (negative) cooperativity. To quantify the cooperativity, we define a collective lowering operator $J = \sum_{i=1}^{N} \sigma_i^-$. Here, the term $\langle J^\dagger J \rangle$ describes the collective behavior of all the emitters in the system, while $\sum_{i=1}^{N} \langle \sigma_i^+ \sigma_i^- \rangle$ only accounts for the sum of the independent behaviors of the individual emitters. From their difference, we evaluate a cooperativity parameter $C = (\langle J^\dagger J \rangle - \sum_{i=1}^{N} \langle \sigma_i^+ \sigma_i^- \rangle)/\langle J^\dagger J \rangle$ and plot it as a function of $P$ at $\kappa = 11.7g$ (Fig. 3g). As expected, the calculated cooperativity shows the same trend as the collective state regimes in Fig. 3e. The existence of a large cavity dissipation ($\kappa \gg N\Gamma$) provides a dominant relaxation channel of the collective system, and therefore, the output field becomes proportional to the degree of collective behavior ($a \propto J$). As a result, the emerged collective coherence is directly reflected in the correlation measurement of emitted photons [29,31], which can be approximated to the form of

$$g^{(2)}(\tau) \approx \frac{\langle J^\dagger(0) J^\dagger(\tau) J(\tau) J(0) \rangle}{\langle J^\dagger(0) J(0) \rangle \langle J^\dagger(\tau) J(\tau) \rangle}. \tag{5}$$

Figure 3f shows the calculated $g^{(2)}(0)$ as a function of $\kappa$ and $P$ in a log scale, while Fig. 3g plots $g^{(2)}(0)$ with $P$ at our experimental condition of $\kappa = 11.7g$. The strong bunching area, corresponding to the steady-state subradiant regime with negative cooperativity, indicates that the bunching value is highly related to $\rho_-$. In Fig. 3h, we plot the $g^{(2)}(0)$ values with $\kappa$ under weak pumping ($P = 9 \times 10^{-4} g$) and observe that giant photon bunching only occurs in the intermediate $\kappa$ regime (maximum at $\kappa/g \approx 3$). In a bad cavity regime with large $\kappa$, the emitters are independently excited to a fully excited state and pass through a collective superradiance state in the decay process, resulting in a low bunching ($g^{(2)}(0) < 2$). This is the typical regime described in Fig. 1b, and most previous superradiance effects have been demonstrated in this regime [2,6,11,36]. In contrast, a high $Q$ cavity with small $\kappa$ stores the coherence in the cavity field rather than in the emitters, consequently exhibiting a lasing feature ($g^{(2)}(0) \approx 1$). Interestingly, the magnitude of the bunching reaches its maximum for two coupled emitters and decreases with the number of emitters, as shown in Fig. 3h. This is because the number of collective channels increases with the number of emitters [29].

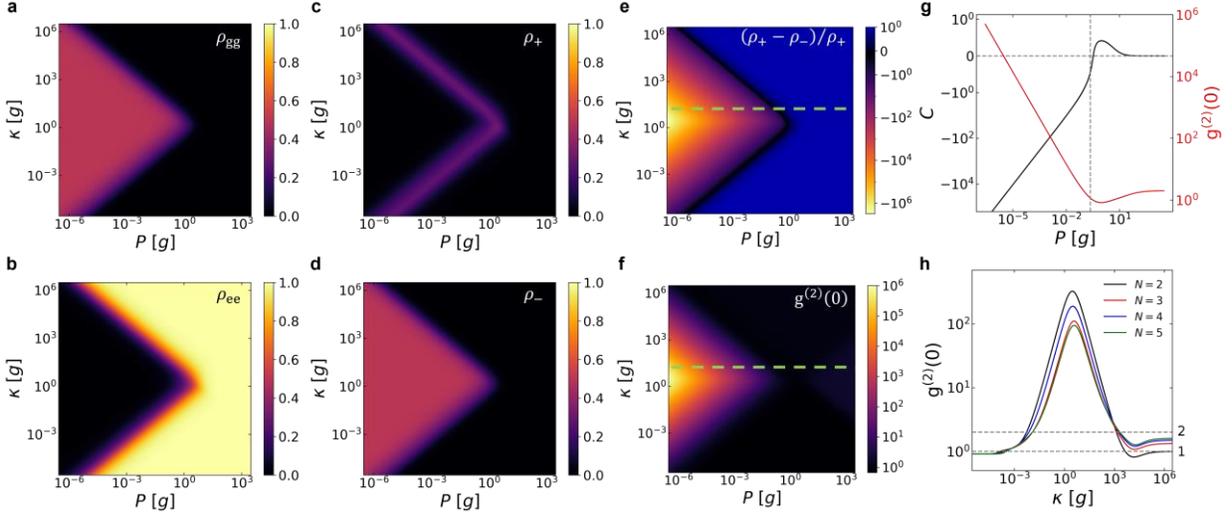

**Fig. 3. Simulation of steady-state population and photon statistics. a-d**, Comparison of steady-state populations of $\rho_{gg}$ (**a**), $\rho_{ee}$ (**b**), $\rho_+$ (**c**), and $\rho_-$ (**d**) with varying $\kappa$ and $P$. **e**, Calculated difference ratio $(\rho_+ - \rho_-)/\rho_+$. The plot is in a symmetric log scale to visualize both positive and negative values. **f**, Calculated second-order coherence function $g^{(2)}(0)$. Green dashed lines in **e** and **f** indicate our experimental condition of $\kappa = 11.7g$. **g**, Cooperativity parameter and $g^{(2)}(0)$ plotted as functions of pumping power $P$ at $\kappa = 11.7g$. **h**, $g^{(2)}(0)$ plotted as a function of $\kappa$ with varying the number of quantum emitters. The maximum of $g^{(2)}(0)$ occurs when $\kappa \approx 3g$ and its magnitude of the peak decreases with the number of quantum emitters in a cavity.

**Long-lived subradiance and giant bunching by suppressed dephasing**

In our experiment shown in Fig. 2d, where a less optimized dissipation rate of $\kappa/g \approx 11.7$ is considered, we observed a rather weak superbunching $g^{(2)}(0) = 3.27$ than the theoretically calculated value of about 54 shown in Fig. 3h. In the above simulation, we assumed ideal quantum emitters without any dephasing for simplicity. However, the QDs suffer from non-negligible dephasing in reality, particularly with the above-band pumping conditions, which diminish the magnitude of cooperativity (Supplementary Section 4). In the experiment, we were able to mitigate the dephasing by controlling the excitation wavelengths[37]. In Figs. 4a-c, we compare the PL spectrum and photon correlation curves under different pumping wavelengths of 790, 870.7, and 891.8 nm, corresponding to above-band, below-band, and quasi-resonant excitation, respectively. The PL spectrum shows that lowering the excitation energy reduces the number of excited QDs until only a single QD peak remains at the excitation wavelength of 891.8 nm. Depending on the excitation wavelength, the correlation data exhibit notable differences. Under above-band pumping (Fig. 4a), a similar bunching characteristic as Fig.2d is measured. However, in Fig. 4b, a significant enhancement in the bunching peak ($g^{(2)}(0) = 8.36 \pm 0.34$) is observed under a below-band pumping condition, which reduces dephasing compared to above-band pumping. This giant bunching is accompanied by an extended antibunching decay time of $36.1 \pm 2.3$ ns, supporting much-suppressed single-photon decay by long-lived subradiance and subsequently more deterministic two-photon pair generation. Another stark change occurs when the excitation energy is further reduced. In quasi-resonant excitation (Fig. 4c), the correlation data suddenly show a distinct antibunching curve ($g^{(2)}(0) = 0.30 \pm 0.003$) with a very fast single-photon decay time of 88.7 ± 5.3 ps. This phenomenon can be attributed to the fact that each QD may have different *p*-shell wavelengths even though they have the same *s*-shell emission wavelength [38]. Therefore, under quasi-resonant excitation of a single QD, the collective two-photon bunching signal disappears at zero time delay. Instead, the single QD in a cavity shows antibunching with a single-photon decay rate over 400 times faster than that of two coupled QDs in a subradiant state. the collective two-photon bunching signal disappears at the zero-time delay, and the single QD exhibits an antibunching signal with a single-photon decay rate that is more than 400 times faster than that of two coupled QDs in a subradiant state in Fig. 4b. The bunching shoulder flanking the antibunching dip in Fig. 4C is attributed to spectral jitter or metastable states, which is well-known in quasi-resonant conditions [39]. Figures 4b and 4c fundamentally demonstrate how a single and two quantum emitters can behave differently in their recombination dynamics and their photon statistics by the collective effects, and it features a strong nonlinearity at a single-emitter level.

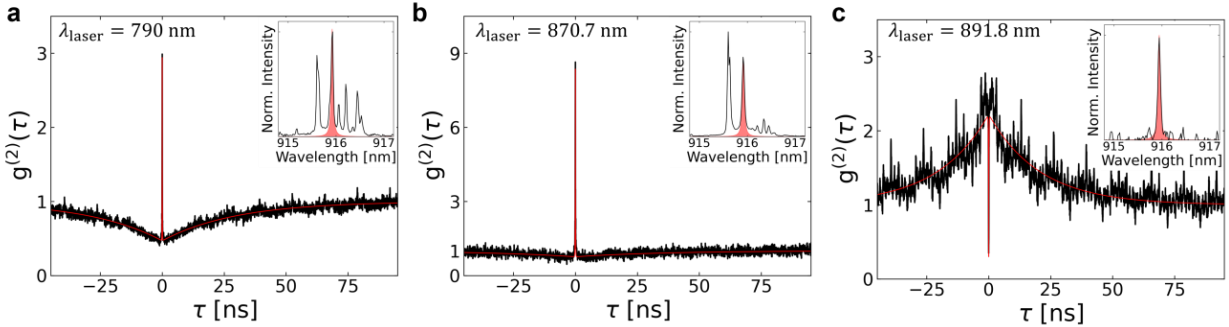

**Fig. 4. Change in photon statistics with excitation wavelength. a-c**, are second-order correlations at different continuous wave excitation wavelengths of 790 nm (**a**) for above-band, 870.7 nm (**b**) for below-band, and 897.8 nm (**c**) for quasi-resonant conditions. In the below band condition (**b**), an enlarged bunching peak ($g^{(2)}(0) = 8.36$) and prolonged antibunching decay (36.1 ns) are observed as a result of suppressed dephasing. The quasi-resonant excitation in **c** selectively excites a single QD, resulting in strong antibunching, $g^{(2)}(0) = 0.30$ at the zero-time delay. Insets display PL spectra at the respective excitation wavelengths. The red color indicates a spectrally filtered QD peak for $g^{(2)}(\tau)$ measurements.

**Conclusion**

We have demonstrated the emergence of a long-lived collective state among multiple emitters through cavity-mediated subradiance. Without the cooperativity between the emitters, steady-state subradiance is only observable in very low pumping power regions, making it have been unexplored experimentally thus far. In this work, QDs are coupled through a balanced dissipation channel and are spatially/spectrally well coupled via a highly directional, low *Q* cavity, which consequently induces negative cooperativity and effectively extends the steady-state subradiance region. As a result, we have observed nonclassical giant bunching, a hallmark of steady-state subradiance. Therefore, our approach reveals that a tailored photonic environment can mediate coherent interactions between quantum emitters and populate them in a long-lived entangled state. Unlike earlier studies that relied on precise coherent control in the pulsed regime, our advancement of steady-state subradiance offers valuable quantum resources such as highly squeezed quantum light and robust-to-dephasing quantum entanglement, suitable for quantum metrologies [17] and quantum storage [15,16]. Our model also demonstrates that maximum entanglement can be achieved with just two quantum emitters in a cavity. Such two-coupled quantum emitters, with a collective decay time much longer than those of the single emitters, could also facilitate the creation of large-scale photonic cluster states [40]. Furthermore, this long-lived collective state has potential applications in energy harvesting [41]. Therefore, our successful demonstration of steady-state subradiance marks a significant advancement toward creating long-lived entanglement and opens new avenues for quantum applications based on highly correlated multi-emitter systems.

## Methods

### Sample design and preparation

The cavity design was optimized for maximizing collection efficiency using a finite-difference time-domain method at a target wavelength of 905 nm, and final parameters are $\Lambda$=321 nm (hole distance along radial direction), R=1.3$\Lambda$ (center disk radius), w=0.43$\Lambda$ (hole distance along tangential direction), and r=0.15$\Lambda$ (hole radius). The calculated cavity mode has a $Q$ value of 1,900 and a Purcell factor of 115 in the simulation.

Self-assembled InAs QDs embedded in 160 nm GaAs on a 1 μm AlGaAs sacrificial layer were grown by molecular beam epitaxy method. Using an electron beam lithography technique, we patterned a structure mask on a bare sample with a 90 nm $Si_3N_4$ hard mask and dry etched it through a reactive ion etching process. The selective wet etching of the sacrificial layer was performed using diluted hydrofluoric acid for a few minutes to create a suspended membrane structure.

### Optical experiment setup

Supplementary Section 5 shows our experimental setup. The sample was cooled down up to 7 K in a closed-cycle helium flow cryostat. Using a polarized beam splitter and two linear polarizers, we enhanced the laser polarization of extinction ratio by more than 1,010:1 to separate a scattered laser from the sample signals. An objective lens of NA 0.7 focused the excitation laser and collected the correlated photons. The photons were coupled with a single-mode fiber using an aspheric lens and sent to a spectrometer with a spectral resolution of 7 GHz to measure a spectrum. For the Hanbury Brown and Twiss measurement, we spectrally filtered the correlated photons using a grating-based bandpass filter with a 36 GHz transmission window and divided two paths using a beam splitter. A superconducting nanowire single photon detector with 35 ps temporal resolution detected photons, and a time-correlated single photon counting with 10 ps temporal resolution recorded the photon correlation.

**Acknowledgments**

The authors acknowledge financial support from the National Research Foundation of Korea grant funded by MSIT (2022R1A2C2003176, RS-2024-00438839, RS-2024-00442762), and Institute of Information & communications Technology Planning & Evaluation (IITP) Grant (RS-2024-00338878, RS-2023-00222863). We acknowledge support from KIST institutional program (2E32942) and UCRF fabrication facilities funded by IITP grant (RS-2023-00227854).

**Author contributions**

K.-Y.K. and W.B.J. designed the sample structure. K.-Y.K. and J.H.L. fabricated the sample. K.-Y.K. and D.H.P. performed the optical experiments. S.I.P. and J.D.S. grew the sample wafer. K.-Y.K., C.L. and J.-H.K. developed the theoretical model and analyzed the data. K.-Y.K. performed the numerical simulation. K.-Y.K. and J.-H.K. prepared the figures. K.-Y.K., C.L., and J.-H.K. wrote the manuscript with input from all authors. C.L. and J.-H.K. supervised the project.

**Competing interests**

Authors declare that they have no competing interests.

**Additional information**

**Supplementary information** is available for this paper.

**Correspondence and requests for materials** should be addressed to Changhyoup Lee or Je-Hyung Kim.

Supplementary Information for

# Long-lived quantum correlation by cavity-mediated subradiance

Kyu-Young Kim[1], Jin Hee Lee[1], Woong Bae Jeon[1], Dong Hyun Park[1], Suk In Park[2], Jin Dong Song[2], Changhyoup Lee[3]*, and Je-Hyung Kim[1]*

[1]Department of Physics, Ulsan National Institute of Science and Technology; Ulsan, 44919, Republic of Korea
[2]Center for Opto-Electronic Materials and Devices Research, Korea Institute of Science and Technology; Seoul, 02792, Republic of Korea
[3]Korea Research Institute of Standards and Science; Dajeon, 34113, Republic of Korea

### 1. Sample optical characterization

To calculate cavity loss $\kappa$, we measured the cavity reflected spectrum. We illuminated a broadband light source to the cavity with weak power and then measured the reflected broadband spectrum. From the Gaussian fitting, we got $1.55 \pm 0.06$ nm FWHM of the cavity.

To confirm the photon statistics of a single emitter, we measured a spectrum of a single QD from the same sample and a second-order correlation function of $g^{(2)}(0) = 0.089$ (Figs. S1).

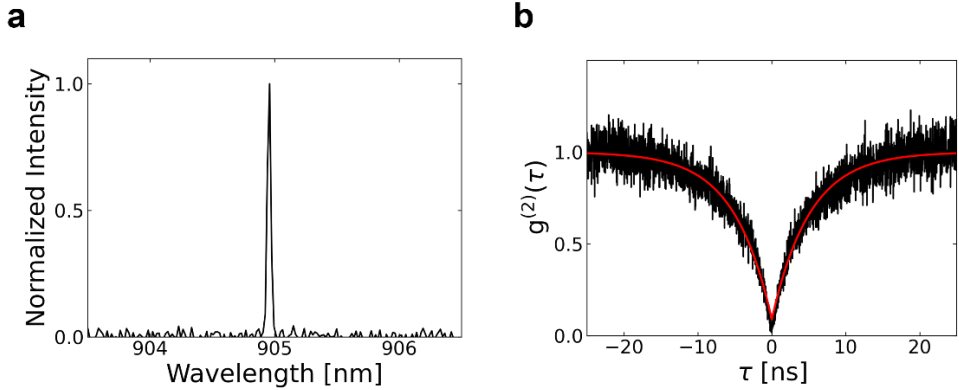

**Fig. S1. Single QD spectrum and the photon statistics. a**, PL spectrum of a single QD. **b**, A second-order correlation functions of the single QD.

To get the lifetime of the cavity-coupled QDs, we proceeded with the time-resolved photoluminescence. We used a Ti:Sapphire pulsed laser of 3 ps pulse width with a 76 MHz repetition rate to excite the emitters. Fig. S2 compares the spontaneous decays of the cavity-coupled QDs and a cavity-uncoupled single QD in an unpatterned bulk sample. We observed 121 ps fast decay time for the cavity-coupled QDs, corresponding to a Purcell enhancement factor of 10.7, compared with the 1.30 ns average lifetime of the cavity-uncoupled single QDs.

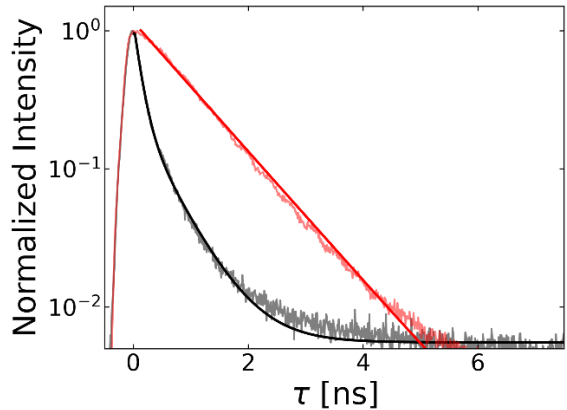

**Fig. S2. Purcell effect in a cavity-coupled single QD.** A comparison of time-resolved photoluminescent curves. A black line represents a decay curve of a cavity-coupled QDs with a decay time of 121 ps. A red line shows a 1.32 ns-long decay time for a single QD in the unpatterned bulk sample.

## 2. Time-dependent collective state population at different $P$

To confirm the dynamics of the collective states over time, we calculate time-dependent populations at different incoherent pumping rates when $\kappa/g = 3$ and compare them with steady-state limits. Figures S3a-c represent time-dependent populations of states, $\rho_{gg}$, $\rho_{ee}$, $\rho_+$, and $\rho_-$ at different pumping rates $P = 0.1, 1, 10g$, respectively. We put the initial population to $|gg\rangle$. In Fig. S3a, the presence of incoherent pumping decreases the ground state population and starts to increase and saturate the populations of other states with time. The saturated time and level of each state largely depend on $P$. In a weak pumping regime, population saturation takes a long time, and $\rho_{gg}$ and $\rho_-$ become equally dominant steady-state populations. As we increase $P$, $\rho_+$ and $\rho_{ee}$ also start to increase. Increasing $P$ further over the collective decay rate, the strong pumping enhances accumulation on $|ee\rangle$ in a short time scale, and $\rho_+$ and $\rho_-$ decrease as $\rho_{ee}$ increases. These results show an influence of $P$ on the state population dynamics and how the states converge to the steady-state limits with time.

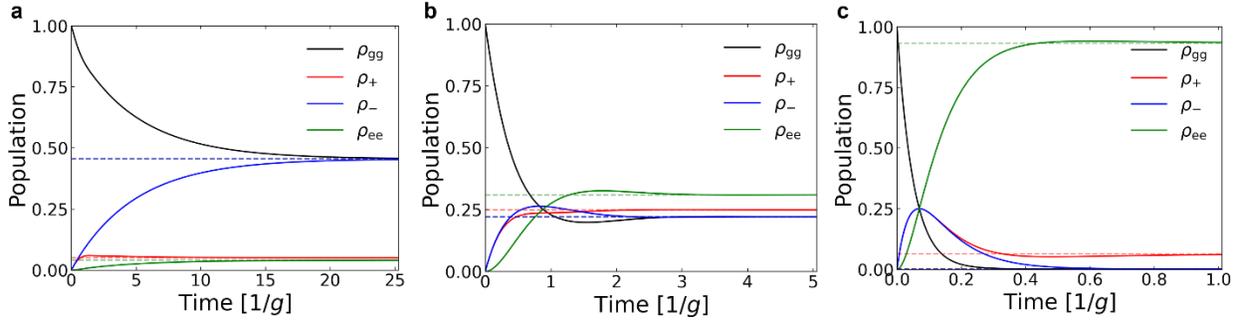

**Fig. S3. Time-dependent populations. a-c,** Solid lines represent time-dependent populations when $\kappa/g = 3$ and $P = 0.1g$ (**a**), $1g$ (**b**), $10g$ (**c**), respectively. Dashed lines are the steady-state limit of each population.

## 3. Coupling strength dependency on the collective state population

The relative ratio between cavity dissipation and coupling strength plays an important role in the steady-state populations of collective states. To find such a balanced regime, we investigate steady-state populations varying with $\kappa$ and the coupling strength independently. To distinguish the coupling strength as a varying variable in this section from $g = 47.4$ GHz, which is used as a unit of rates, we denote the varying coupling strength as $g'$. Figures S4a-d represent steady-state populations at a fixed $P = 9 \times 10^{-4} g$. When $P \ll \kappa, g'$, as increasing $\kappa$, the cavity mode field leaks into free space before the cavity-mediated interaction of emitters, and the steady-state easily accumulates on $|ee\rangle$. However, the large $g'$ compensates for such dissipation loss and delays the point where $\rho_{ee}$ becomes dominant. Therefore, $\kappa$ and $g'$ present a linear relationship for achieving a dominant steady-state population on $|-\rangle$. However, as the pumping rate exceeds the cavity dissipative rate and the coupling strength ($P \gg \kappa, g'$), $\rho_{ee}$ is dominant regardless of $\kappa$ and $g'$ due to strong incoherent pumping. Such a relation is also found in $(\rho_+ - \rho_-)/\rho_+$ (Fig. S4e) and $g^{(2)}(0)$ (Fig. S4f). Therefore, we can find a balanced $\kappa/g' \approx 3$ for achieving steady-state subradiance and maximum $g^{(2)}(0)$.

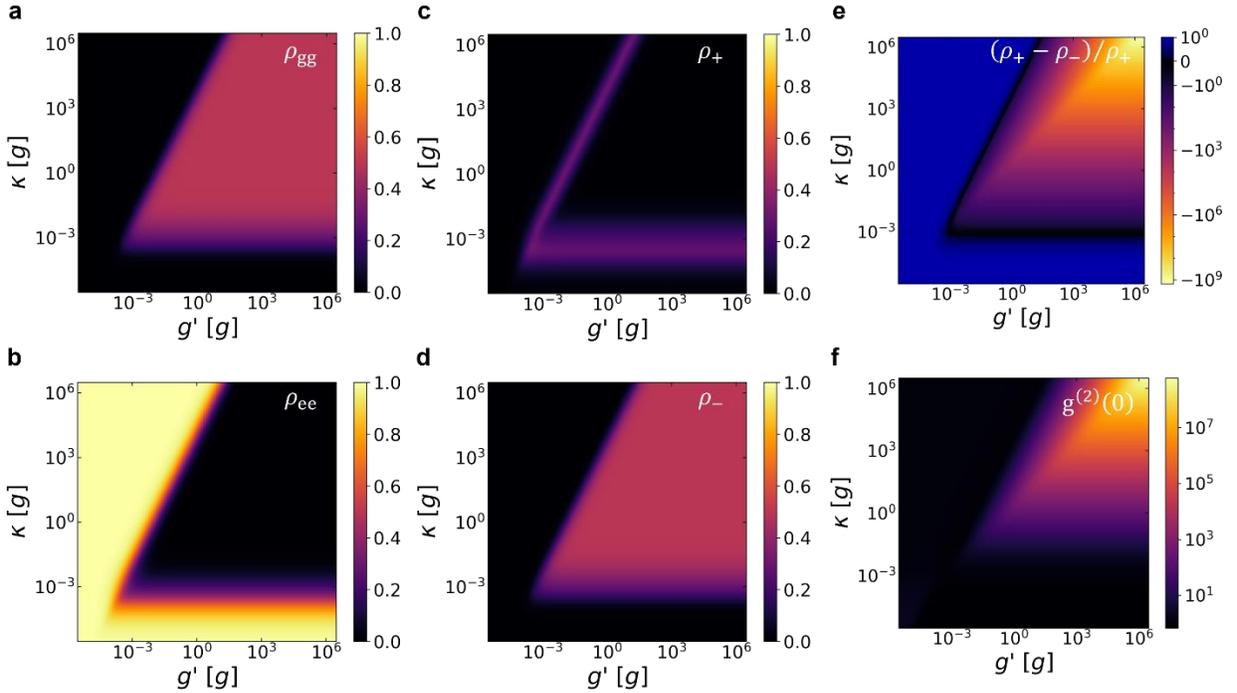

**Fig. S4. Balanced cavity dissipation rate and coupling strength. a-d**, Mapping data of expectation values of collective states of $\rho_{gg}$ (**a**), $\rho_{ee}$ (**b**), $\rho_+$ (**c**), and $\rho_-$ (**d**) as varying $\kappa$ and the coupling strength $g'$ when $P = 9 \times 10^{-4} g$. **e**, Calculated $(\rho_+ - \rho_-)/\rho_+$ in a symmetric log scale. **f,** Calculated second-order correlation at zero-time delay in a log scale.

## 4. Dephasing dependency on the collective state population

To analyze a relation between $\gamma$ and the photon statistics, we calculate the populations and the second-order correlation at the zero-time delay. As we mentioned in the main text, the dephasing causes a flip between subradiant and superradiant states. Therefore, when $P < \gamma$ and $\gamma < \kappa$, the subradiant state flips to superradiant state and goes to the ground state within $\Gamma$. This process makes the ground state population larger than the subradiant state population. Since both the subradiant and superradiant states are excited to the fully excited state by the incoherent pumping at large $P$, the $P$ where the fully excited state population is dominant does not change much. However, as increasing $\gamma$ more than $\kappa$, large $\gamma$ disturbs the cavity-emitters interaction and the QDs work as independent emitters. So, the steady-state population goes to the fully excited state. As a result, calculated $(\rho_+ - \rho_-)/\rho_+$ (Fig. S5a) and $g^{(2)}(0)$ (Fig. S5b) varying with $P$ and $\gamma$ show that emitted photons dramatically lose their correlation as increasing $\gamma$, but the balanced $\kappa/g \approx 3$ still holds (Fig. S6).

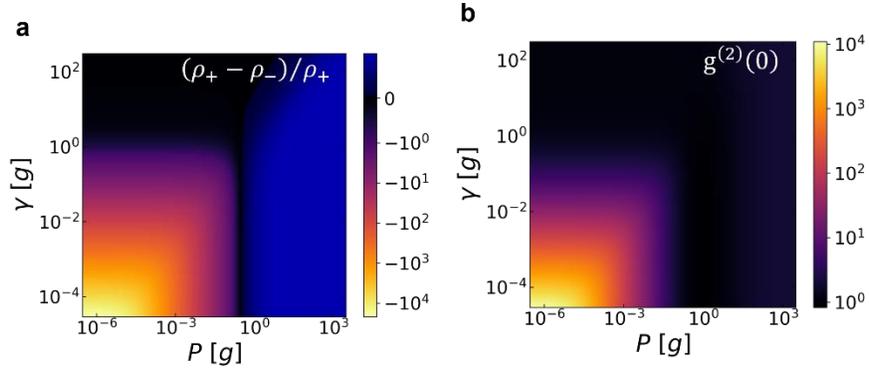

**Fig. S5. Influence of dephasing on steady-state population and second-order correlation. a**, Calculated $(\rho_+ - \rho_-)/\rho_+$ in a symmetric log scale when $\kappa = 11.7g$. **b**, Calculated the second-order correlation at zero-time delay in a log scale.

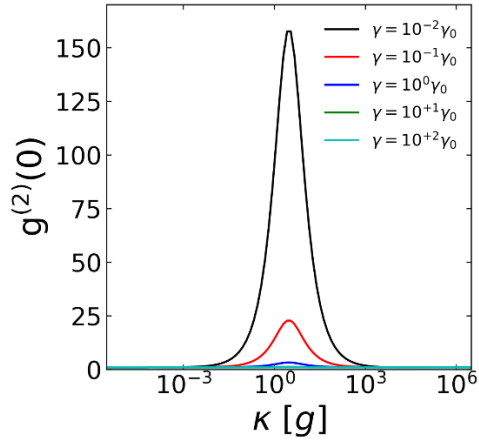

**Fig. S6. Simulated $g^{(0)}(0)$ values as a function of $\kappa$ with the presence of different amounts of dephasing $\gamma$.**

## 5. Experimental Setup schematic

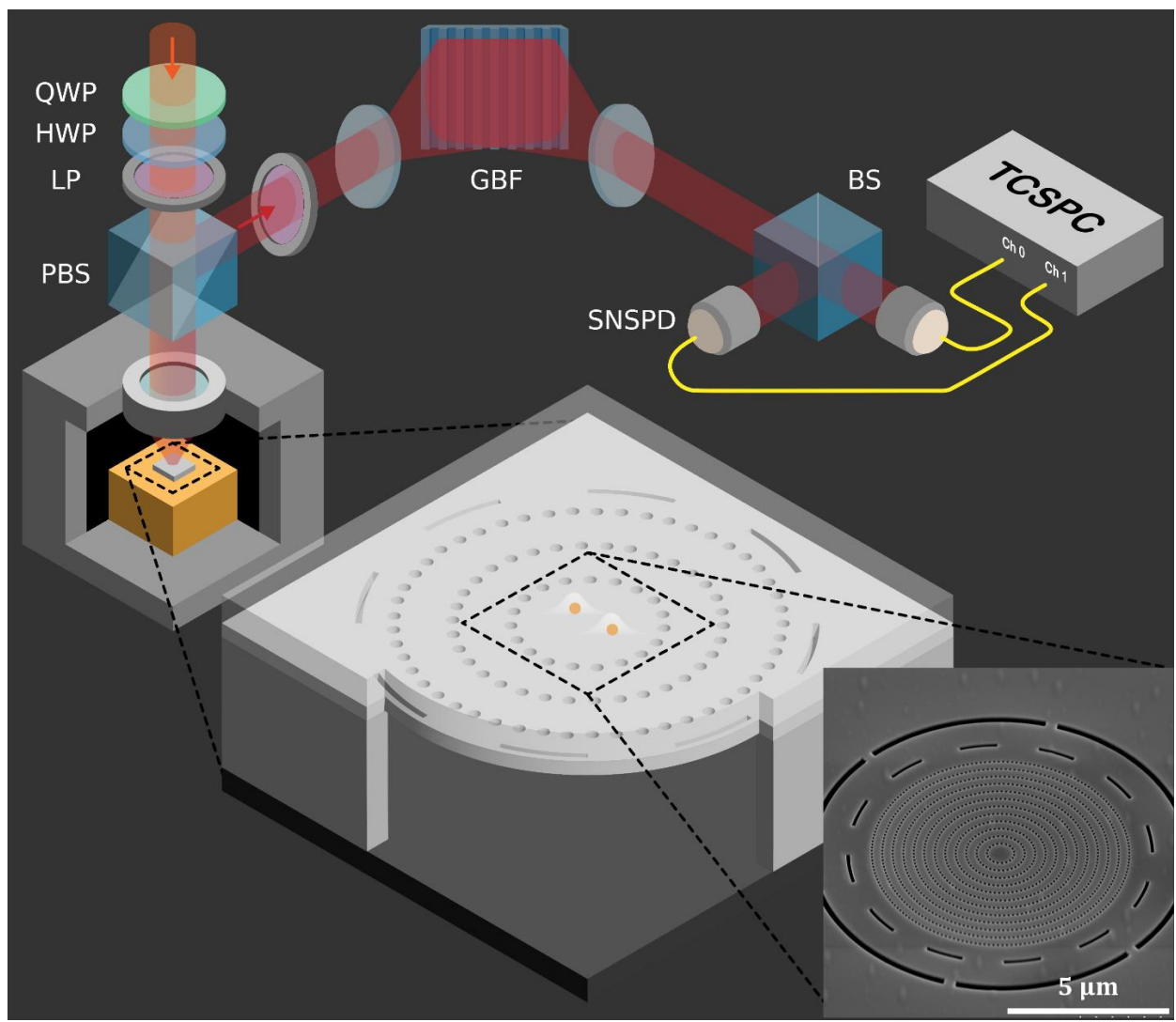

**Fig. S7. Schematic of experimental setup.** Inset is a scanning electron microscopy image of the 45˚ tilted sample.

6